\pdfoutput=1

\documentclass[10pt,conference]{IEEEtran}

\usepackage[T1]{fontenc}
\usepackage[utf8]{inputenc}
\usepackage{dirtytalk}
\usepackage[inline]{enumitem}
\usepackage{listings}
\usepackage[hyphens]{url}

\newcommand{\inlinecode}[1]{\texttt{#1}}
\newcommand{\mem}{memory}
\newcommand{\os}{operating system}

\makeatletter

\def\lst@makecaption{
  \def\@captype{table}
  \@makecaption
}

\makeatother
\makeatletter

\begin{document}

\title{Who Ate My Memory?\\Towards Attribution in Memory Management}

\author{
    \IEEEauthorblockN{Gunnar Kudrjavets, Ayushi Rastogi}
    \IEEEauthorblockA{
        \textit{University of Groningen}\\
                9712 CP Groningen, Netherlands \\
                g.kudrjavets@rug.nl, a.rastogi@rug.nl}

    \and

    \IEEEauthorblockN{Jeff Thomas, Nachiappan Nagappan}
    \IEEEauthorblockA{
        \textit{Meta Platforms, Inc.} \\
                Menlo Park, CA 94025, USA \\
                jeffdthomas@meta.com, nnachi@meta.com}
}

\maketitle

\IEEEtriggeratref{16}

\begin{abstract}
To understand applications' \mem\ usage details, engineers use instrumented builds and profiling tools.
Both approaches are impractical for use in production environments or deployed mobile applications.
As a result, developers can gather only high-level memory-related statistics for deployed software.
In our experience, the lack of granular field data makes fixing performance and reliability-related defects complex and time-consuming.
The software industry needs lightweight solutions to collect detailed data about applications' \mem\ usage to increase developer productivity.
Current research into \mem\ attribution-related data structures, techniques, and tools is in the early stages and enables several new research avenues.
\end{abstract}


\section{Background and motivation}

This paper is motivated by our industry experience while working on problems related to optimizing and tracking \mem\ usage in both commercial and open-source software products.
We observe the presence of a homogeneous set of issues related to \emph{attribution} (\say{How much \mem\ is used by a particular component?}) and \emph{accountability} (\say{What component is exceeding their \mem\ budget?}).
Answering those questions correctly and continuously is necessary to ensure the application's performance and reliability.

Software \emph{performance} is associated with user satisfaction and how actively users engage with an application~\cite{hort_2021}.
One criterion that impacts an application's performance is its \emph{\mem\ usage}~\cite{gregg_systems_2020}.
Characteristics such as the total size of allocations, allocation rate, or the allocated \mem\ type serve as metrics describing how efficiently an application uses \mem.
Memory is one of the most precious system resources on mobile platforms that do not support paging, such as {iOS}~\cite{levin_ios_2017}.
A suggested design pattern for environments where \mem\ is limited is to make each component responsible for accounting for its own \mem\ usage~\cite{noble_2002}.

A detailed understanding of \mem\ usage is also required to increase \emph{reliability}.
A variety of modern kernels and \os s based on them, such as {Android}, {FreeBSD}, {iOS}, or {Linux}, use a system component called an \emph{out-of-\mem\ killer}~\cite{android_oomk,fb_oomk,linux_oomk}.
An out-of-\mem\ killer is responsible for terminating processes when their \mem\ usage exceeds a certain quota, or excessive \mem\ usage puts the stability of an entire \os\ in danger.
\emph{Engineers need to understand the application's memory footprint in detail to avoid premature termination}.

In industry, we note that organizations develop their in-house ecosystems to solve company-specific \mem\ attribution issues.
A similar trend is present with open-source software where projects such as {M}ozilla {F}irefox have a product-specific extensive framework to track and attribute \mem\ usage~\cite{firefox_memory}.
We observe the shortcomings in \mem\ attribution capabilities and tools even in popular kernels such as {L}inux that has three decades of real-world usage and an active developer community~\cite{lwn_quote}.

Most of the research related to \mem\ management focuses on increasing the performance of custom allocators~\cite{leijen2019mimalloc,jansson_rpmalloc,snmalloc,lee_2014,evans_2006} or ensuring their correctness~\cite{ball_2001,appel_2020}.
The limited existing research into \mem\ attribution is
in the early stages,
involves invasive profiling techniques,
and is specific to the {M}essage {P}assing {I}nterface~\cite{gutierrez_2020}.

\section{State of affairs}

\begin{flushright}
\emph{\say{\dots how it is that memory can be allocated without the kernel knowing where it went.
The problem is that the tracking infrastructure just isn't there.}}
\end{flushright}
\begin{flushright}
--- 2022 Linux Storage, Filesystem, MM and BPF Summit
\end{flushright}

\medskip

Modern \os s enable tracking \mem\ usage at the process level~\cite{tanenbaum_modern_2001,russinovich_windows_2012,love_linux_2005,stallings_operating_2009}.
However, the \emph{granularity at the process level is insufficient for engineers to perform efficient debugging or performance engineering}~\cite{gregg_systems_2020}.
A common task that engineers encounter in practice is \emph{determining what portion of the allocated amount of \mem\ is consumed by a specific component, scenario, or a subsystem}.
In the context of performance or reliability engineering that can mean a dynamic library, feature, function, or thread.
Data about \mem\ consumption is necessary to determine
\begin{enumerate*}[label=(\alph*),before=\unskip{ }, itemjoin={{, }}, itemjoin*={{, and }}]
    \item what components to optimize
    \item if performance regressions are present
    \item what performance tuning techniques to use.
\end{enumerate*}
It is also necessary to understand how \mem\ usage changes over time.
For example, after a new commit, updates to the toolset, such as a compiler, change in dependent libraries, or change in some application's configuration parameters.

The standard approach to acquiring details about an application's \mem\ usage is to use a \emph{profiler}.
When an application is executed under the profiler, such as {V}isual {S}tudio {P}rofiler~\cite{vs_profiler} or {X}code~\cite{xcode_profiler}, then the profiler tracks each allocation and its source.
An engineer can later filter, query, and visualize the resulting dataset.
For example, the profiler may record the complete stack information coupled with the allocation size and originating dynamic library.
Gathering the profiler data is time-consuming, can require the application to be compiled with specific flags (in case an instrumentation framework such as {V}algrind~\cite{valgrind} is used), and requires a significant amount of disk storage to store the entire history of allocations.
Above all, \emph{this approach is not practical outside the application's development environment}.
The overhead caused by instrumentation can cause an application to execute an order of magnitude slower.

\subsection{Existing attribution techniques}

Two main approaches to attribute \mem\ usage in source code exist: annotating each call to allocate \mem\ and annotating each scope.
We describe the existing usage patterns for each technique.

In kernel mode, Windows uses \emph{pool tagging}~\cite{russinovich_windows_2012}.
Each driver can specify a pool tag when it requests to allocate \mem~\cite{Oney1999},~\cite{Dekker1999},~\cite{Orwick2007}.
Depending on a size of a driver, it can use one or more pool tags to differentiate between various subsystems.
Pool tags also have another function in Windows---kernel crashes if the driver does not release all the allocations with a specific tag when the driver is unloaded~\cite{russinovich_windows_2012}.

On mac{OS}, a caller can use a function such as \inlinecode{OSMalloc} and associate each allocation with an opaque tag.
However, that tag is used only for reference counting~\cite{singh_mac_2016}.
The tag count is increased by one each time a specific tag is allocated.
In user mode, macOS also enables passing custom tags generated by \inlinecode{VM\_MAKE\_TAG} macro to functions such as \inlinecode{vm\_allocate}~\cite{singh_mac_2016}.
Listing~\ref{code:tagged_alloc} shows a sample usage pattern when a custom tag is associated with an allocation.

\lstset{language=C,breaklines=true,frame=single,basicstyle=\ttfamily\small,caption={Tagged allocation in mac{OS}.},label=code:tagged_alloc}
\begin{lstlisting}
/* An allocation billed to networking. */
err = vm_allocate(..., VM_MEMORY_LIBNETWORK);
\end{lstlisting}

\smallskip

One of the {FreeBSD} ports is a basic heap \mem\ accounting system \inlinecode{libpdel}~\cite{libpdel} that similarly requires each caller to specify a \say{\mem\ type} in a form of a string.

As a result of annotations, the application can during the runtime enumerate its virtual \mem, and gather the distribution and size of allocations per a different \mem\ tag or a type.

\emph{Hierarchically} tracking memory usage by annotating source code is another option.
Developers need to attribute each scope with a specific tag.
All the allocator activity in that scope and its children will be attributed to that tag.
Listing~\ref{code:hierarchical_alloc} displays how all the allocations in the function \inlinecode{bar()} and its children will be \say{billed} to the tag \inlinecode{foo} using the example of \inlinecode{TfMallocTag} tagging system~\cite{TfMallocTag}.

\lstset{language=C,breaklines=true,frame=single,basicstyle=\ttfamily\small,caption={Allocation tracking using the TfMallocTag system.},label=code:hierarchical_alloc}
\begin{lstlisting}
void bar() {
    TfAutoMallocTag tag("foo");

    funcA();
    funcB();
}
\end{lstlisting}

\subsection{Limitations of current techniques}

All these approaches have constraints because they require
\begin{enumerate*}[label=(\alph*),before=\unskip{ }, itemjoin={{, }}, itemjoin*={{, and }}]
    \item annotation of each allocation or scope
    \item complete source code to be available
    \item hierarchical tracking needs to intercept all the allocations in the current process.
\end{enumerate*}
However, a standard application has dependencies, such as system or third-party libraries.
Without modifying the dependencies, the ability to track allocations in detail is limited to the application's \say{own code.}

We are unaware of any \os s, languages, or tools that enable engineers to query and keep track of \mem\ attribution at a granular level \emph{without sacrificing the application's performance}.
The exploration of possible solutions is still in the early stages.
The most recent proposal for \mem\ allocation tracking in {Linux} is from August 2022~\cite{linux_mem_tracking}.
The lack of these facilities negatively impacts each non-trivial software project.
Understanding the application's \mem\ usage in detail
is realistic only in the development environment.
However, in our experience, \emph{predicting or debugging an application's behavior in the production environment based on the data from the development environment is ineffective}.

\section{Future research directions}

The choice of the abstraction layers and a variety in the problem space enables multiple research avenues.
Ideally, the support for \mem\ attribution will be integrated throughout the \os, \mem\ allocator, and a runtime library such as the {GNU} {C} Library (glibc)~\cite{gnuc}.
Support for programming languages used for systems programming (e.g., {C}, {C++}, {Rust}) is imperative.

The desired solution to improve engineers' ability to attribute \mem
\begin{enumerate*}[label=(\alph*),before=\unskip{ }, itemjoin={{, }}, itemjoin*={{, and }}]
    \item can be enabled and disabled on demand
    \item has a minor performance overhead and will be usable in the production environment
    \item enables querying the memory attribution during runtime
    \item has a well-designed set of \textsc{API}s.
\end{enumerate*}

One potential practical approach we envision in user mode is the \emph{usage of custom \mem\ allocators~\cite{evans_2006,evans_2015,evans_scalable_2011,lee_2014} to assist with the attribution}.
Custom allocators such as jemalloc intercept each allocation request made in the context of a process.
Therefore, the intercept mechanism can track all the metadata such as allocation size, current thread, timestamp, or specific flags passed to the function.
The intercept mechanism can use either data from the application that specifies the current attribution scope, classify callers based on sampling the call stack, or use some other techniques.

\bibliographystyle{IEEEtran}
\bibliography{memory-tracking}

\end{document}